\documentclass[notitlepage,12pt,english]{revtex4}
\usepackage[T1]{fontenc}
\usepackage[latin9]{inputenc}
\usepackage[a4paper]{geometry}
\usepackage{amsmath}
\usepackage{graphicx}
\usepackage{amssymb}
\usepackage{esint}

\usepackage{float}

\makeatletter
\@ifundefined{textcolor}{}
{%
 \definecolor{BLACK}{gray}{0}
 \definecolor{WHITE}{gray}{1}
 \definecolor{RED}{rgb}{1,0,0}
 \definecolor{GREEN}{rgb}{0,1,0}
 \definecolor{BLUE}{rgb}{0,0,1}
 \definecolor{CYAN}{cmyk}{1,0,0,0}
 \definecolor{MAGENTA}{cmyk}{0,1,0,0}
 \definecolor{YELLOW}{cmyk}{0,0,1,0}
 }

\makeatother

\usepackage{babel}

\begin{document}

\title{Spontaneous symmetry breaking and phase coherence of a photon Bose-Einstein condensate coupled to a reservoir}

\author{Julian Schmitt, Tobias Damm, David Dung, Christian Wahl, Frank Vewinger, Jan Klaers\footnote{Present address: Institute for Quantum Electronics, ETH Z\"urich, Auguste-Piccard-Hof 1, 8093 Z\"urich, Switzerland}, Martin Weitz}
\affiliation{Institut f\"ur Angewandte Physik, Universit\"at Bonn, Wegelerstr.
8, 53115 Bonn, Germany}

\begin{abstract}

\noindent\\
We examine the phase evolution of a Bose-Einstein condensate of photons generated in a dye microcavity by temporal interference with a phase reference. The photo-excitable dye molecules constitute a reservoir of variable size for the condensate particles, allowing for grand canonical statistics with photon bunching, as in a lamp-type source. We directly observe phase jumps of the condensate associated with the large statistical number fluctuations and find a separation of correlation timescales. For large systems, our data reveals phase coherence and a spontaneously broken symmetry, despite the statistical fluctuations.

\end{abstract}

\maketitle
\noindent\\
At the heart of Bose-Einstein condensation, the phase transition of a cold and dense gas of integer spin (bosonic) particles to a macroscopically populated ground state, is its phase coherence~\cite{Pitaevskii,Anderson}. While for a thermal, incoherent ensemble each particle evolves individually, in a Bose-Einstein condensate the macroscopic ground state occupation leads to the whole condensate acting as a single, giant quantum wave. Each individual measurement will then yield a fixed, though random phase, as expected from spontaneous symmetry breaking~\cite{Javanainen,Naraschewski,Castin}. The macroscopic phase of Bose-Einstein condensates has been verified in interference experiments with ultracold atomic gases~\cite{Andrews}. For condensates of polaritons, mixed states of matter and light, polarization symmetry breaking was reported~\cite{Baumberg,Ohadi}, while polariton arrays have shown phase locking~\cite{Anton}. Further, spatial coherence has been reported for equilibrium photon condensates~\cite{Klaers4}, and also in a nonequilibrium regime~\cite{Marelic2}. Both atomic and polariton condensates are usually assumed to emit a single wave train of constant amplitude~\cite{Pitaevskii,Deng}. They operate with an essentially fixed number of particles, corresponding to a description in the microcanonical or canonical ensemble limit. Such sources have both first and second-order coherence.

The predictions by statistical physics dramatically change for Bose-Einstein condensates that are subject to particle (and heat) exchange with a reservoir. In the grand canonical statistical ensemble, the population of each energy state of the Bose gas performs uncorrelated number fluctuations of order of its mean occupation number~\cite{Huang}. When applied to the macroscopically occupied ground state of a Bose-Einstein condensate, this implies large statistical number fluctuations of order of the total particle number occurring deep in the condensed phase, a behavior termed the \glqq grand canonical fluctuation catastrophe\grqq~\cite{Ziff,Holthaus,Kocharovsky,Fierz,Fujiwara}. Experimentally, grand canonical conditions in the condensed phase have been realized in a photon Bose-Einstein condensate coupled to a dye medium~\cite{Klaers2,Marelic}, where thermalization of the photon gas is achieved by absorption and re-emission processes on dye molecules~\cite{Klaers3,Kirton2,Schmitt2}. The photo-excitable dye molecules here do not only act as a heat reservoir for the photon gas, but also as a particle reservoir, thus the number of condensate particles can fluctuate around a mean value. For large reservoir sizes, strong photon number fluctuations in the condensed phase, as predicted by statistical theory~\cite{Klaers1,Sobyanin}, have been experimentally verified in the dye microcavity system~\cite{Schmitt1}. Condensates in the grand canonical ensemble limit have a vanishing second-order coherence, corresponding to a zero-delay intensity correlation $g^{(2)}(0)=2$, same as incoherent sources, as lamps, have~\cite{Loudon}. Correspondingly, the question arises whether the macroscopic ground state occupation leads to phase coherence despite large statistical fluctuations. 

The essential physics of a grand canonical source in the condensed phase can be modeled by a wave train of variable amplitude, see Fig.~1(a), subject to statistical amplitude fluctuations due to the interconversion between condensate particles and dye electronic excitations. In general, the photon statistics interpolates between Poissonian statistics for small reservoir sizes and a Bose-Einstein distribution for an infinitely large reservoir~\cite{Klaers1,Schmitt1}. While for Poissonian statistics damped intensity fluctuations and macroscopic phase coherence of the condensate are expected~\cite{Snoke,DeLeeuw1,Kirton1,DeLeeuw3}, for the latter distribution the fluctuations become as large as the average photon number $\bar n$, i.e. $\Delta n = \bar n$. Therefore, both in the grand canonical and intermediate statistical regime there is a finite probability $P_0$ that the cavity contains no photons at all, which causes the condensate population $n(t)$ to occasionally drop to zero intensity. When the cavity now resumes emission, as indicated in the right panel of Fig.~1(a), the condensate phase will be lost.

In this Letter, we experimentally examine the temporal phase evolution of a Bose-Einstein condensate of photons realized in a dye microcavity by beating its emission with the output of a narrowband laser source. While a phase-stable interference signal is monitored in the canonical statistical regime of the photon condensate, when tuning towards the grand canonical limit intensity fluctuations lead to phase jumps in the interference signal. The measured oscillation resumes with a random phase, as a consequence of symmetry breaking following spontaneous emission. The observed phase jump rate scales linearly with the inverse system size, suggesting full first-order coherence in the limit of large condensates despite photon bunching, i.e. $g^{(2)}(0)>1$, and we observe separate timescales of first and second-order coherence. Our findings reveal realization of an optical source with unusual coherence properties.

\begin{figure}[t]
\begin{centering}
\vspace{1mm}
\includegraphics[width=8.2cm]{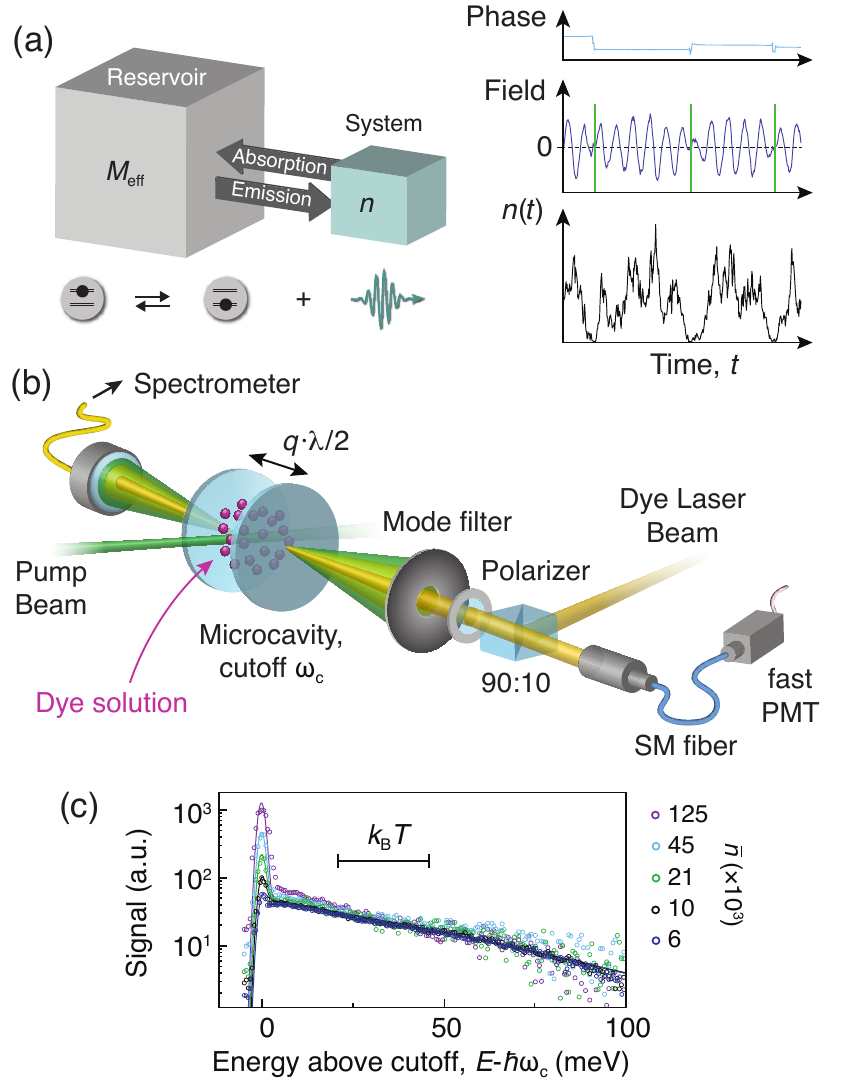}
\par\end{centering}
\vspace{-1mm}
\caption{(a) Representation of the statistical system (left), where dye molecules act as heat bath and particle reservoir for the photon gas. When the effective reservoir size $M_\textrm{eff}$ is large, grand canonical statistical conditions are fulfilled and the corresponding emission (right) exhibits photon bunching and random phase jumps following intensity drops. (b) Overview of the experimental setup. From the emission out of the dye-filled microresonator the condensate mode is filtered, and after a polarizer overlapped with the laser reference. The resulting beat signal is detected on a photomultiplier tube. Simultaneously, radiation transmitted through the second cavity mirror at the reverse side is used to record spectra of the photon gas. (c) Experimental (circles) and theory (lines) spectra of the cavity emission show the saturation of thermal modes at the onset of condensation and, with the known critical photon number at condensation threshold, allow to determine the condensate mode population $\bar n$. (Cutoff energy $\hbar\omega_\textrm{c} = 2.13~\textrm{eV}$)}
\vspace{-2mm}
\end{figure}

A schematic of the apparatus used to generate Bose-Einstein condensation of a two-dimensional photon gas, which has been described in detail in Ref.~\cite{Klaers4}, is shown in Fig.~1(b). Our experiment confines photons in a dye-solution filled microcavity made of two curved mirrors spaced by $1.4~\mu\textrm{m}$~\cite{Klaers2,Supplemental}. The small cavity length constitutes an energy separation of longitudinal modes comparable to the spectral width of the dye emission, which reduces the photon dynamics to the transverse motional degrees of freedom with the longitudinal wave number ($q{=}7$) remaining fixed. Effectively, this introduces a low-energy cutoff of $\hbar\omega_\textrm{c}$, corresponding to a wavelength typically selected to be in the range of $560\textrm{-}610~\textrm{nm}$, and imprints a spectrum of cavity photon energies restricted to well above the thermal energy with non vanishing chemical potential, see Fig.~1(c). To both introduce an initial cavity photon population and compensate for losses, the dye is pumped with an external laser beam. By repeated absorption re-emission processes, the two transversal cavity modal quantum numbers thermalize to the (rovibrational) dye temperature $T=300~\textrm{K}$, while the thermalization process conserves the longitudinal mode number. When spectrally monitoring the cavity photons, we observe a thermal distribution of photon energies above the low-energy cutoff, which extends over a range of more than $3 k_\textrm{\tiny B}T$, see Fig.~1(c). Despite pumping and losses, the photon gas is well described by an equilibrium distribution as thermalization by dye absorption and re-emission occurs faster than the timescale at which a photon is lost~\cite{Klaers3,Kirton2,Schmitt2}. The photon gas inside the resonator has a quadratic dispersion and is formally equivalent to a two-dimensional gas of harmonically confined massive bosons~\cite{Supplemental,Klaers3}, such that Bose-Einstein condensation is expected to occur~\cite{Bagnato}. When increasing the total cavity photon number above a critical value ($N_\textrm{c}\simeq 80,000$ for the used parameters), in addition to the thermal cloud a macroscopic population of the cavity ground mode is observed, see Fig.~1(c) for spectra of the Bose-Einstein condensed photon gas~\cite{Klaers2,Marelic}. In the course of thermalization photons are frequently converted into dye electronic excitations and vice versa, and the dye acts both as a particle reservoir and a heat bath for the photon gas (Fig.~1(a), left)~\cite{Klaers1,Schmitt1}. This situation allows the photon number in the cavity ground mode to fluctuate around the average value $\bar n$ (Fig.~1(a), right). In contrast to earlier studies, we are here interested in the phase evolution of a photon Bose-Einstein condensate in the presence of those statistical number fluctuations. In particular, the relation between first and second-order coherence is studied, with an emphasis on the corresponding timescales and their behavior when approaching the limit of large condensate fractions upon conservation of the fluctuation level $\Delta n/ \bar n= \sqrt{g^{(2)}(0)-1}$.

To measure the condensate phase evolution, radiation transmitted through one cavity mirror is spatially filtered in the far field to suppress thermal modes. The unpolarized condensate emission passes a polarizer and is overlapped with radiation of a dye laser. The dye laser with a linewidth of $250~\textrm{kHz}$ here serves as a stable phase reference, and the phase evolution is encoded in the temporal interference between the light sources. For this, both sources are coupled into a single mode fiber and the resulting beat signal is detected with a fast photomultiplier ($0.6~\textrm{GHz}$ bandwidth) and monitored with an oscilloscope operating in fast frame mode. The resulting interference signal can be written as
\begin{equation}
I(t) = I_\textrm{c} + I_\ell + 2 \sqrt{I_\textrm{c} I_\ell} \cos\left[(\omega_\textrm{c} - \omega_\ell)t + \Delta \phi(t)\right],
\label{eq1}\vspace{-0.5mm}
\end{equation}
\noindent
where $I_\textrm{c}$~($I_\ell$) and $\omega_\textrm{c}$~($\omega_\ell$) denote the intensities and frequencies of the photon condensate (dye laser) beam, respectively, at the fiber output and $\Delta\phi(t)$ is the relative phase. In addition to serving as a local oscillator, the use of an external optical source as phase reference avoids the possible influence of phase locking between nearby placed sources, an effect observed e.g. in polariton arrays~\cite{Anton}. Our detected beat signals, see e.g. Fig.~2(a), show a relatively slow frequency chirp attributed to a density modulation of the dye solution after the onset of the pumping pulse, as detailed in the Supplementary Information~\cite{Supplemental}. The condensate linewidth is thus expected to be composed of contributions from both the frequency drift and a small Schawlow-Townes-like phase diffusion~\cite{DeLeeuw1}, as well as from discrete phase jumps owing to statistical number fluctuations. To obtain the latter contributions, the recorded beat signals are analyzed for irregularities in their oscillation period by an algorithm, which was tested to allow detection of discrete phase jumps with phase rotations in the range of $\Delta\phi=\left[0.2\pi,1.8\pi\right]$. Experimentally, phase jump rates (number of phase jumps per time interval) ranging from $5~\mu\textrm{s}^{-1}$ to $200~\mu\textrm{s}^{-1}$ are resolved.

\begin{figure}[t]
\begin{centering}
\includegraphics[width=8.2cm]{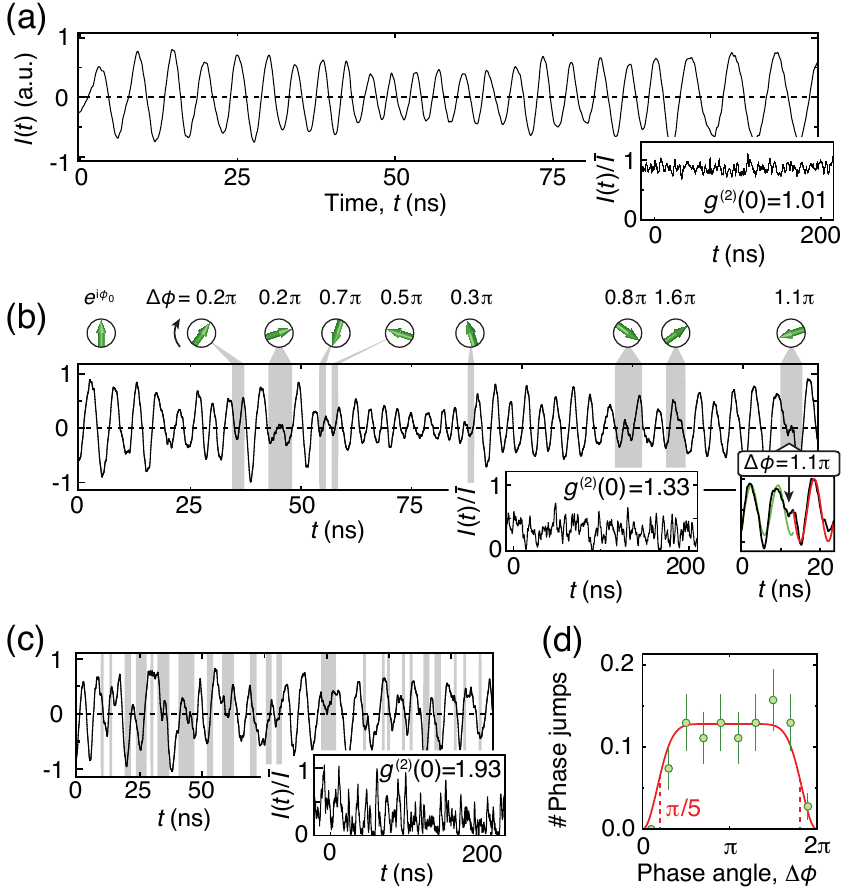}
\par\end{centering}
\vspace{-1mm}
\caption{Temporal interference signals between photon condensate and dye laser, for average condensate photon number of (a) $114,000$, (b) $8,300$, and (c) $3,700$ respectively, realizing different levels of statistical number fluctuations (dye concentration $\rho=3.10^{-3}~\textrm{mol/l}$, cutoff wavelength $\lambda_c=582~\textrm{nm}$), the insets give condensate intensity traces for the corresponding parameters. The shaded areas show phase jumps detected by our algorithm, increasing in rate from (b) to (c). The shown values for the phase angles of the oscillation were obtained by fitting the beat signal in the vicinity of the phase jumps, see the inset on the right hand side of (b) for an example. The smallest phase rotation detectable by our algorithm is $0.2\pi$.~(d) Histogram of observed phase jump angles along with a fit of the detection window (solid line).}

\vspace{-3mm}
\end{figure}

The main plots in Fig.~2 give three different beat signals, recorded with condensates of different levels of the number fluctuation, indicated by $g^{(2)}(0)$, for which different condensate sizes were used. The insets result from a second run with only the dye microcavity emission irradiating on the photomultiplier detector, allowing a determination of the level of intensity fluctuations for a corresponding data set, from which the second-order coherence function $g^{(2)}(\tau)$ is determined. For essentially second-order coherent light (Fig.~2(a), $g^{(2)}(0)=1.01$) we find a beat signal with no observable phase jumps. Figures 2(b) and 2(c) give signals recorded with condensates of larger intensity fluctuations, with $g^{(2)}(0)=1.33$ and $1.93$ respectively, which show an irregular beat signal with clear phase jumps in the oscillation, as indicated by the shaded areas. The associated rotation angles of the condensate phase, indicated by arrows on top of Fig.~2(b), are determined from a fit to the beat signal in the vicinity of phase irregularities, as exemplified in the inset of Fig.~2(b). We attribute the increasing phase jump rate of those condensates to the here enhanced probability for the condensate photon number to drop below that required to maintain phase coherence. On the other hand, the timescale for intensity fluctuations here remains almost constant, indicating a separation of coherence timescales. Fig.~2(d) shows a distribution of observed phase angles, which follows the expected $U(1)$ symmetry within the detection window. Upon re-establishment of the condensate emission the new oscillation resumes with an arbitrary phase.

\begin{figure}[t]
\begin{centering}
\includegraphics[width=8.2cm]{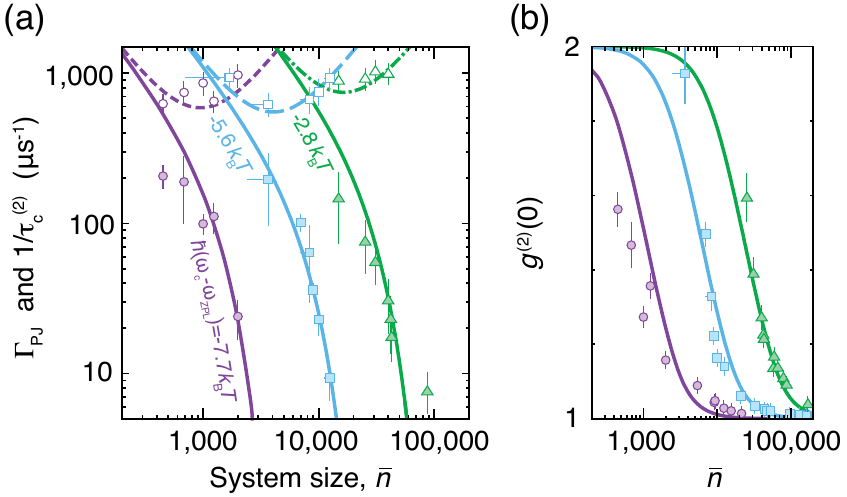}
\par\end{centering}
\vspace{-1mm}
\caption{(a) Separation of timescales for first- and second-order coherence. Inverse second-order coherence times $1/\tau_c^{(2)}$ (open symbols) and phase jump rates $\Gamma_\textrm{PJ}$ (solid symbols) for three different effective reservoirs $M_{\textrm{eff}}$, of relative size $\{1,10,120\}$ indicated by violet circles, blue squares and green triangles, respectively, varied using the dye-cavity detuning $\omega_\textrm{c}-\omega_\textrm{\tiny ZPL}$. For the experimental parameter range, the observed phase jump rates are significantly below the inverse second-order coherence time. (b) Zero-delay autocorrelation $g^{(2)}(0)$ versus condensate population $\bar n$ for the corresponding parameters. For an increased effective reservoir size, photon bunching with $g^{(2)}(0)>1$ is maintained up to larger sizes of the Bose-Einstein condensate and over a broad range coexists with separated values for $\Gamma_\textrm{PJ}$ and $1/\tau_c^{(2)}$. The shown lines in (a) and (b) are theory curves, for $M = \{2.0,4.5,5.0\}\times 10^9, \hat B_{12} = \{140,250,1300\}\textrm{s}^{-1}$. }
\vspace{-2mm}
\end{figure}

For a quantitative analysis of the timescales, the solid symbols in Fig.~3(a) show the rate of the observed phase jumps versus the number of condensate photons for three different effective reservoir sizes, here tuned by varying the detuning between the condensate frequency $\omega_\textrm{c}$ and the dye zero-phonon line at $\omega_\textrm{\tiny ZPL}=2\pi c/(545~\textrm{nm})$. A smaller detuning increases the effective reservoir size 
$M_\textrm{eff} = M/\{2 + 2\cosh\left[\hbar (\omega_\textrm c - \omega_\textrm{\tiny ZPL} )/k_\textrm{\tiny B} T \right]\}$ with $M$ as the number of dye molecules, as it increases the average number of electronically excited dye molecules when the average photon number in the condensate is retained~\cite{Klaers1,Schmitt1}. Correspondingly, we observe the persistence of statistical number fluctuations up to larger condensate populations $\bar n$, see Fig.~3(b). The phase jump rate $\Gamma_\textrm{PJ}$ increases both for larger effective reservoirs and smaller numbers of condensate photons (system size). This is well understood from the in these cases larger probability to reach a very small number of photons, which cannot anymore sustain phase coherence of the condensate. In a heuristic approach, we expect the phase jump rate to follow the rate $\Gamma_\textrm{PJ}^0=P_0 M \hat B_{12}$ for a vanishing photon number in the cavity, where $ \hat B_{12}$ is the Einstein rate coefficient for absorption. The zero-photon probability $P_0$ is obtained from numerical calculations~\cite{Supplemental}, and the corresponding results for $\Gamma_\textrm{PJ}^0$ are shown as solid lines in Fig.~3(a). We are aware that also small, but non-zero, photon numbers can lead to phase jumps. However, the good agreement with the experimental results indicates that $\Gamma_\textrm{PJ}^0$ captures the essential physics. This is supported by our Monte Carlo simulations~\cite{Supplemental}, which are also in agreement with the scaling for the condensate linewidth predicted for the grand canonical limit~\cite{Kirton1}. For comparison, the open symbols in Fig.~3(a) show the inverse of the observed second-order coherence times along with theory (dashed lines)~\cite{Supplemental}. Clearly, a separation between the shown timescales characterizing first and second-order coherence, respectively, is observed in the presence of strong photon bunching. Though our analysis does not capture diffusive linewidth contributions, our finding is in clear contrast to predictions for chaotic light, for which the two timescales are identical~\cite{Loudon}. This shows that despite the large statistical number fluctuations a spontaneously broken symmetry exists.

\begin{figure}[t]
\begin{centering}
\includegraphics[width=8.2cm]{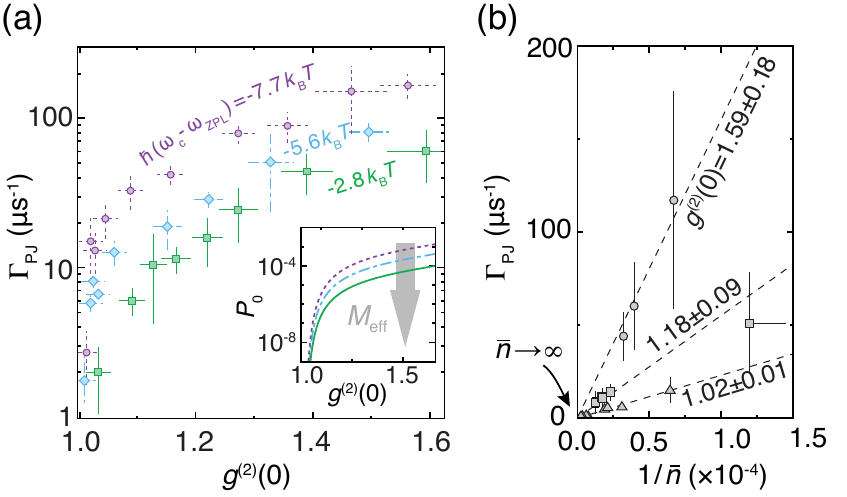}
\par\end{centering}
\vspace{-1mm}
\caption{Coherence properties in the limit of large systems. (a) Phase jump rate as a function of the zero-delay autocorrelation $g^{(2)}(0)$ for three different values of the dye-cavity detuning $\omega_\textrm{c}-\omega_\textrm{\tiny ZPL}$, which tunes the effective reservoir size $M_\textrm{eff}$. The inset shows the calculated value of the zero-photon probability $P_0$ for the corresponding parameters. (b) Phase jump rate versus the inverse condensate size $1/\bar n$ for data sets with three different levels of intensity fluctuations (symbols), indicated by the corresponding $g^{(2)}(0)$. A linear extrapolation $\bar n\rightarrow \infty$ for all data sets (dashed lines) suggests full phase coherence in this limit despite the persistence of photon bunching.}
\vspace{-2mm}
\end{figure}

We next address the question of the phase stability of a flickering condensate expected in the limit of an infinitely large photon number ($\bar n\rightarrow \infty$). To reach this limit, we here besides the condensate size additionally increase the reservoir size such that the fluctuation level remains fixed, i.e. $g^{(2)}(0)~=~\textrm{const.}$, which from theory  is expected for a fixed ratio $\bar n^2/M_\textrm{eff}$~\cite{Klaers1,Supplemental}. Figure 4(a) shows the variation of the phase jump rate on the measured zero-delay correlation function $g^{(2)}(0)$ for three different reservoirs. From this data we extract the phase jump rate as a function of the inverse condensate size ($1/\bar n$) for three fixed values of the measured zero-delay second-order coherence function $g^{(2)}(0)$, see Fig.~4(b). The data points for the extrapolation all are in a regime with $\bar n^2\geq M_\textrm{eff}$ where a separation of timescales for $\Gamma_\textrm{PJ}$ and $1/\tau_\textrm{c}^{(2)}$ was observed (Fig.~3(a)). We point out that in the far grand canonical case $\bar n^2\ll M_\textrm{eff}$, for which $g^{(2)}(0)$ approaches $2$ and $\Gamma_\textrm{PJ} \simeq 1/\tau_\textrm{c}^{(2)}$, we also in the thermodynamic limit do not expect a separation of timescales. For the shown data, see Fig.~4(b), a linear extrapolation to an infinitely large condensate ($1/\bar n\rightarrow 0$) indicates a vanishing phase jump rate for all three shown data sets, with $g^{(2)}(0)=1.59(18), 1.18(9)$ and $1.02(1)$, respectively. Even if macroscopic phase coherence is never lost, amplitude modulation of the oscillator will lower the degree of the first-order coherence and result in a finite linewidth also in the limit of an infinitely large system.

To conclude, we have observed spontaneous symmetry breaking in an optical condensate with bunched number statistics. Our data reveals a separation of characteristic timescales for first and second-order coherence properties. From a thermodynamic viewpoint, the observed regime with well-defined phase of the grand canonical statistics condensate means that an order parameter exists despite the large number fluctuations~\cite{Zannetti}.

The results are expected to have direct implications for studies of the Josephson effect and photonic lattices building upon condensates coupled to both heat and effective particle reservoirs, as realizable in the dye microcavity system~\cite{Hartmann,Greentree,Angelakis,Schiro,Umucalilar,DeLeeuw2}. From a technical perspective, speckle-free and other optical imaging can benefit from grand canonical statistics light sources in the condensed phase~\cite{Redding}.

We thank Y. Castin for his generalized calculation of the second-order coherence time. This work has been financially supported by the DFG (We1748-17) and the ERC (INPEC).

\end{document}